\pdfoutput=1
\documentclass[journal=apchd5,manuscript=letter]{achemso} %,layout=twocolumn
\setkeys{acs}{articletitle = true}

\usepackage{xcolor}
\usepackage{hyperref}
\usepackage{graphicx}
\usepackage{siunitx}
\usepackage[utf8]{inputenc}
\usepackage{textcomp}
\hypersetup{colorlinks,linkcolor={blue},citecolor={blue},urlcolor={red}}

\newcommand{\VSi}{V$_\textnormal{Si}^{-}$}

\newcommand{\BO}{$B_0$}

\newcommand{\pmSTH}{$\pm$3/2}
\newcommand{\pmSOH}{$\pm$1/2}
\newcommand{\pSTH}{$+$3/2}
\newcommand{\pSOH}{$+$1/2}
\newcommand{\mSTH}{$-$3/2}
\newcommand{\mSOH}{$-$1/2}
\newcommand{\sipm}[3]{#1 $\pm$ \SI{#2}{#3}}

\DeclareSIUnit\gauss{G}

\title{Coherent electrical readout of defect spins in 4H-SiC  by photo-ionization at ambient conditions}

\author{Matthias Niethammer}
\email{matthias.niethammer@pi3.uni-stuttgart.de}
\author{Matthias Widmann}
\author{Torsten Rendler}
\author{Naoya Morioka}
\author{Yu-Chen Chen}
\author{Rainer St\"ohr}
\affiliation{3rd Institute of Physics and Center for Applied Quantum Technologies, University of Stuttgart, 70569 Stuttgart, Germany}
\author{Jawad Ul Hassan}
\affiliation{Department of Physic, Chemistry and Biology, Link\"oping University, SE-581 83 Link\"oping, Sweden}

\author{Shinobu Onoda}
\author{Takeshi Ohshima}
\affiliation{National Institutes for Quantum and Radiological Science and Technology, Takasaki 370-1292, Japan}

\author{Sang-Yun Lee}
\affiliation{Center for Quantum Information, Korea Institute of Science and Technology, Seoul 02792, Republic of Korea}
\author{Amlan Mukherjee}
\affiliation{3rd Institute of Physics and Center for Applied Quantum Technologies, University of Stuttgart, 70569 Stuttgart, Germany}
\author{Junichi Isoya}
\affiliation{Faculty of Pure and Applied Sciences, University of Tsukuba, Tsukuba 305-8573, Japan}
\author{Nguyen Tien Son}
\affiliation{Department of Physic, Chemistry and Biology, Link\"oping University, SE-581 83 Link\"oping, Sweden}
\author{J\"org Wrachtrup}
\affiliation{3rd Institute of Physics and Center for Applied Quantum Technologies, University of Stuttgart, 70569 Stuttgart, Germany}
\alsoaffiliation{Max Planck Institute for Solid State Research, 70569 Stuttgart, Germany}

\date{\today}
\DeclareUnicodeCharacter{2212}{-}
\DeclareSIUnit{\belmilliwatt}{Bm}
\DeclareSIUnit{\dBm}{\deci\belmilliwatt}
\begin{document}

\maketitle

\begin{abstract}
Quantum technology relies on proper hardware, enabling coherent quantum state control as well as efficient quantum state readout.
In this regard, wide-bandgap semiconductors are an emerging material platform with scalable wafer fabrication methods, hosting several promising spin-active point defects.
Conventional readout protocols for such defect spins rely on fluorescence detection and are limited by a low photon collection efficiency.
Here, we demonstrate a photo-electrical detection technique for electron spins of silicon vacancy ensembles in the 4H polytype of silicon carbide (SiC).
Further, we show coherent spin state control, proving that this electrical readout technique enables detection of coherent spin motion.
Our readout works at ambient conditions, while other electrical readout approaches are often limited to low temperatures or high magnetic fields.
Considering the excellent maturity of SiC electronics with the outstanding coherence properties of SiC defects the approach presented here holds promises for scalability of future SiC quantum devices.
\end{abstract}
\section{Keywords}
silicon vacancy center, silicon carbide, PDMR, ODMR, electrical readout, coherent control

\section{Main text}
Solid state color centers have developed into a leading contender in quantum technology  owing to their vast potential as hardware for quantum sensing and quantum networks~\cite{degen_quantum_2017, casola_probing_2018,atature_material_2018, yang_high-fidelity_2016, nagy_high-fidelity_2018,humphreys_deterministic_2018}. Typically, these techniques employ optical control for spin state initialization and readout.
Spins in solids can provide long spin relaxation and dephasing times and therefore constitute excellent quantum bits.
In certain cases, \latin{e.g.} for spins in wide-bandgap semiconductors, single spin manipulation and optical spin state readout is feasible~\cite{gruber_scanning_1997, jelezko_observation_2004, widmann_coherent_2015, christle_isolated_2015}.
A number of systems, like spin dopants in silicon~\cite{morello_single-shot_2010, morton_embracing_2011,zwanenburg_silicon_2013}
 or quantum dots~\cite{hanson_single-shot_2005,godfrin_electrical_2017}, allow for electrical spin readout.
However, because their spin polarization typically relies on Boltzmann statistics, they require low temperature operation or large magnetic fields~\cite{van_tol_high-field_2009}.

In contrast, color centers in wide-bandgap semiconductors show efficient optical spin polarization at room temperature~\cite{waldherr_dark_2011,tetienne_magnetic-field-dependent_2012}. Electrical readout of color center spins at ambient conditions relies on an efficient mechanism for spin-to-current conversion.
This can be realized by measuring a laser induced spin-dependent photocurrent, which is often referred to as photocurrent detected magnetic resonance (PDMR).
Several publications  have successfully demonstrated this principle for various materials~\cite{stegner_electrical_2006,mccamey_spin_2008, franke_spin-dependent_2014,lee_spin-dependent_2010}.
Recently, this technique has been applied to the nitrogen-vacancy (NV) center in diamond, by combining electrical readout with optical excitation~\cite{bourgeois_photoelectric_2015,hrubesch_efficient_2017} and even achieved single defect~\cite{siyushev_photoelectrical_2019} detection.
It turns out that the signal-to-noise ratio (SNR) in this approach is competitive to optical detection~\cite{siyushev_photoelectrical_2019} and at the same time allows better integration into electronic periphery.
However, diamond as host material is not compatible with industrial technologies, \latin{e.g.} large-scale wafers and the development of efficient diamond electronics is still subject to research.
Silicon carbide (SiC) on the other hand has attracted attention due to its outstanding optical, electrical and mechanical properties~\cite{atature_material_2018}.

Traditionally, interest in defects in SiC was driven by their impeding properties to high power electronic devices~\cite{kimoto_fundamentals_2014}.
This has initialized a wealth of studies utilizing electron paramagnetic resonance~\cite{isoya_epr_2008, mizuochi_continuous-wave_2002} and electrically detected magnetic resonance~\cite{cochrane_edmr_2010,sato_electrically_2006,gruber_electrically_2018, umeda_electrically_2012,cottom_recombination_2016}.
Among many investigated phenomena, spin dependent recombination has been shown to allow for self-calibrating magnetometers in a non-coherent fashion~\cite{cochrane_vectorized_2016}.
In addition, several spin-active defects with long spin coherence times~\cite{simin_locking_2017,christle_isolated_2015} even at room temperature~\cite{widmann_coherent_2015,carter_spin_2015} have been found.
The quantum properties of such color centers have lately been used to demonstrate magnetic field and temperature sensing~\cite{simin_high-precision_2015, simin_all-optical_2016, niethammer_vector_2016, anisimov_optical_2016}.
In this work, we demonstrate electrical readout of a negatively charged silicon-vacancy (\VSi{}) spin ensemble in a 4H-SiC device via PDMR at ambient conditions.

The negatively charged silicon vacancy \VSi{} at the cubic lattice site (V$_\textnormal{2}$) in 4H-SiC provides both, stable deep level energy states in a wide-bandgap host and a spin dependent intersystem crossing (ISC).
Previous studies revealed, that the defect has a spin quartet  manifold of S=3/2~\cite{wimbauer_negatively_1997,mizuochi_continuous-wave_2002} in ground state (GS) and excited state (ES), which are separated by \SI{1.35}{\electronvolt} (\SI{916}{\nano\meter})~\cite{sorman_silicon_2000}. GS and ES Landé g-factors are identical (g=2.003) and their respective zero field splittings (ZFS) are \SI{70}{\mega\hertz} and $\approx$\SI{410}{\mega\hertz}~\cite{tarasenko_spin_2018} at ambient conditions.
In addition, a long-lived metastable state gives rise to non-radiative and spin-dependent ISC relaxation, enabling optical spin state initialization and readout under ambient conditions
\cite{sorman_silicon_2000,kraus_room-temperature_2014,  widmann_coherent_2015}. 
Furthermore, it provides excellent coherence times even at room-temperature \cite{widmann_coherent_2015, carter_spin_2015, yang_electron_2014}.

\begin{figure}
\includegraphics[width=1\linewidth]{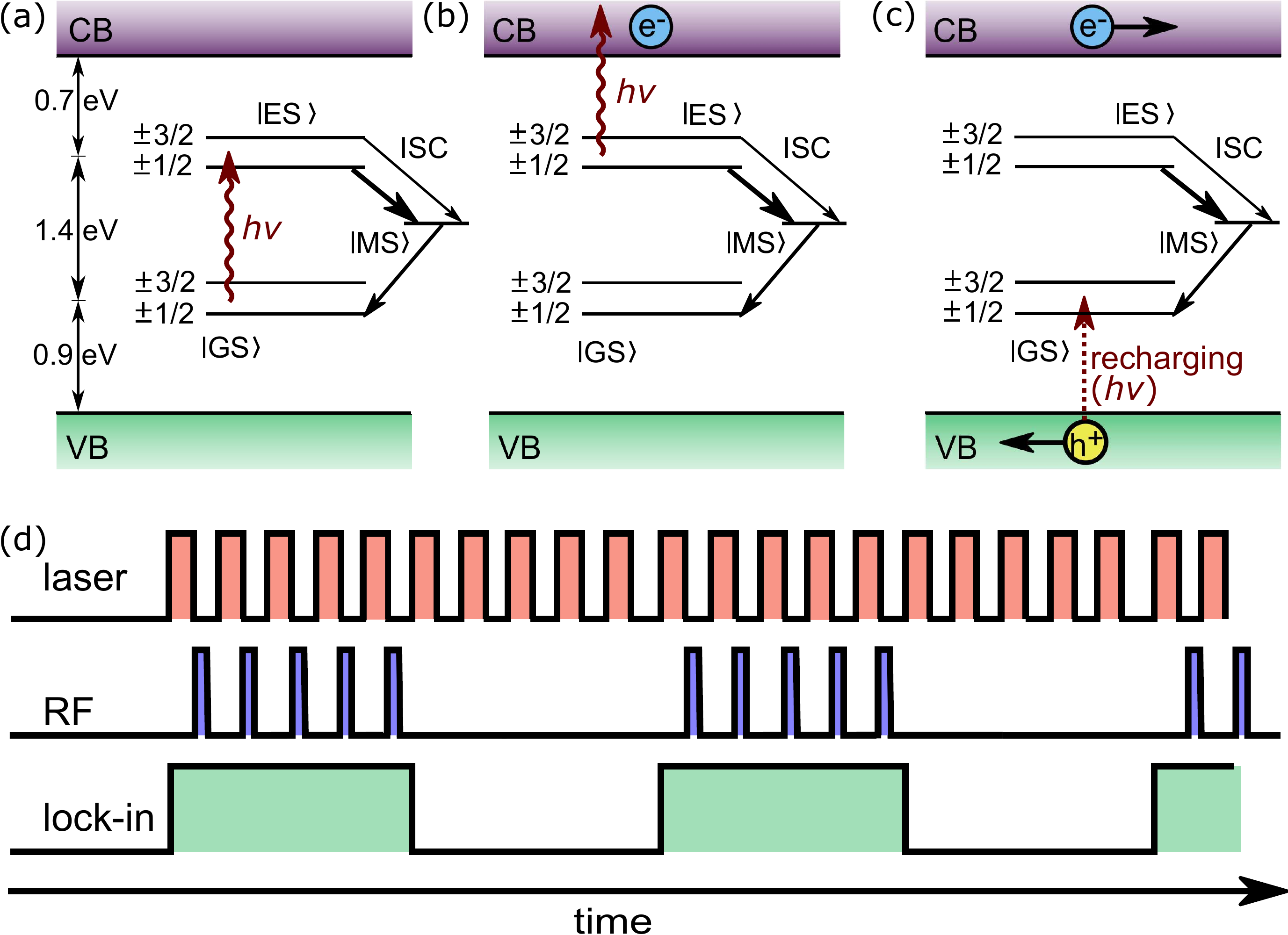}
\caption{\label{fig:pulses}PDMR mechanism and readout. (a)-(c) Spin-dependent photo-ionization. (a) Single photon excitation. ES can relax to either GS or MS. ISC to MS is dependent on the spin state in ES, thus GS is polarized. (b) Second photon ionizes the defect and introduces a free electron in the CB. (c) Recharging of the defect from VB and separation of charges lead to a photocurrent. (d) Pulse sequence scheme used for lock-in detection. See text for detail.} 
\end{figure}

In the following, we discuss the principle of PDMR and how it can be applied to \VSi{}.
Figs.~\ref{fig:pulses}(a)-(c) depict the underlying charge dynamics:
a deep level defect absorbs a photon and is promoted from its GS to the ES.
From there:
(i) The system can decay back to the GS by emitting a photon.
(ii) The system can undergo a non-radiative ISC \latin{via} a metastable state (MS). A spin-state dependency of this ISC rate is usually exploited in optically detected magnetic resonance (ODMR).
(iii) While being in the ES, the system can undergo a second optical excitation to the conduction band (CB).
In case (iii), an excess electron populates the CB, and the defect charge state $n$ is changed to $n+1$.
To reach a steady-state charge distribution, the defect can re-capture an electron either from the CB, or from other recharging sources, \latin{e.g.} from other defects in the surrounding, or from the valance band (VB) through photo-induced electron-hole pair generation.
In the third case, the free electron in the CB and the hole in the VB can be measured as photocurrent.

If the ISC rates are spin-dependent, this charge circulation enables photo-electrical spin-state readout.
Note that the second photon may also be absorbed by the MS during an ISC cycle. 
Because the overall lifetime in the ES and MS is determined by the ISC as well, the spin dependency of the ISC rate alters the chance for the second photon absorption. 
The amount of spin-dependent contribution to photocurrent by this process is then expected to be the sum of currents created by promoting an electron either from the ES or MS to the CB.

We assume the \VSi{} to be initialized in the \pmSOH{} spin subspace of the GS by optical illumination.
During optical excitation, the ES is populated.
If the ISC rate from ES to MS states is higher for \pmSOH{} than for the \pmSTH{} states,
the chance for two-photon absorption from the ES of \pmSTH{} states is higher.

Populating the \pmSTH{} states by resonantly driving the spin transition will consequently increase the photocurrent. 
For an ionization from the MS to CB, a decrease in current should be measured. 
The overall sign and magnitude of the effect will thus be determined by the difference in absorption cross section, ISC rates, lifetime and population of the ES and MS.

\begin{figure}
\includegraphics[width=1\linewidth]{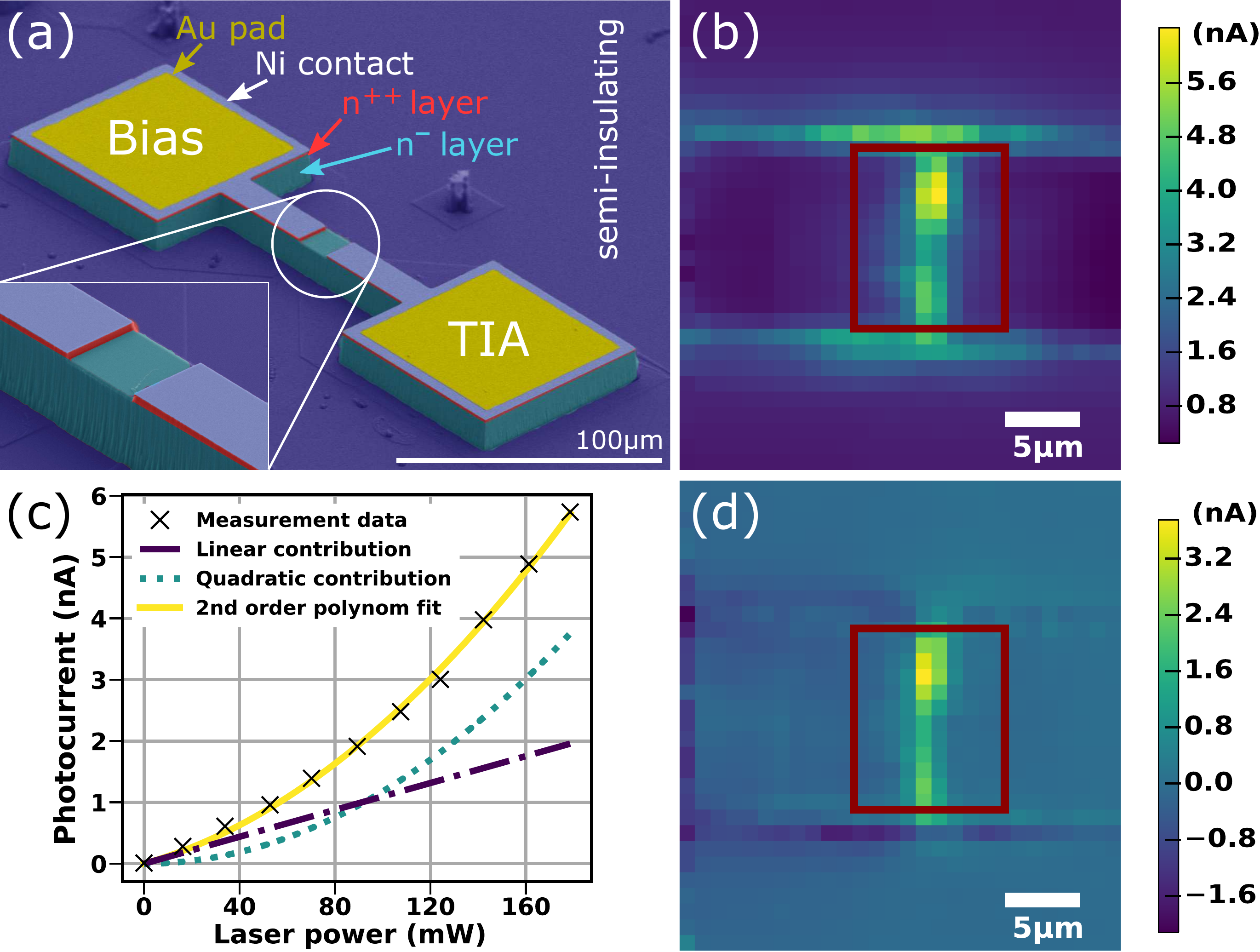}
\caption{\label{fig:sample} (a) SEM picture (functional layers false color coded) of a fabricated device. Bias and transimpedance amplifier (TIA) connections are marked. Inset: Zoom-in of the etched optical opening.  (b) Photocurrent map at -\SI{10}{\volt} bias. Approximate position of optical access opening marked in red. (c) Laser power dependence of the photocurrent at the position of maximum two-photon contribution corresponding to a bright spot in (d). Fit $f(x)=ax^2+bx$ separates linear and quadratic contributions, where $x$ stands for optical power. Fit parameters are: $a$=\sipm{117.77}{6.43}{\femto \ampere}/\si{\milli\watt}$^2$, $b$=\sipm{10.96}{0.93}{\pico\ampere}/\si{\milli\watt}. (d) Map of two-photon excitation contribution to photocurrent at maximum laser power obtained \latin{via} fit parameter $a$.}
\end{figure}

The microstructure used in this work is a n$^{++}$/n$^{-}$/n$^{++}$ metal-semiconductor-metal (MSM) junction, which is shown in Fig.~\ref{fig:sample}(a).
Starting from a n-type 4H-SiC substrate, epitaxial growth was used to fabricate a three-layer stack: 
(i) a \SI{10}{\micro\meter}-thick vanadium-doped semi-insulating layer to reduce leakage currents into the substrate,
(ii) a \SI{10}{\micro\meter}-thick n$^{-}$ layer with N-doping concentration of \SI[mode=math]{1e14}{\centi\meter ^{-3}}, and
(iii) a \SI{400}{\nano\meter}-thick n$^{++}$ layer with N-doping concentration of \SI[mode=math]{8e17}{\centi\meter ^{-3}}.
A nickel (Ni) layer of \SI{100}{\nano\meter} thickness was deposited forming a Schottky contact on the n$^{++}$ layer.
The sample was etched down by \SI{10}{\micro\meter}, leaving fingers of various width as devices.
Subsequently, the Ni and n$^{++}$ films were removed in rectangular center areas of various sizes, for optical access to the n$^{-}$ region (see zoomed inset in Fig.~\ref{fig:sample}(a)).
In this layer, we expect the charge state of the \VSi{} to be stable.
Additionally, gold is deposited on the contact pads for wire bonding (see Supporting Information).
After recording $I$--$V$ curves of the device, we create a \VSi{} ensemble  by electron irradiation at \SI{2}{\mega\electronvolt} with a dose of \SI[mode=math]{1e17}{\centi\meter ^{-2}}.
This process degrades the contact quality and device conductivity due to carrier compensation~\cite{ohyama_radiation_2005} (see Supporting Information).
However, this also results in minimizing the dark current, enabling us to maximize the amplifier gain, which is beneficial for electrical readout.
We chose to perform measurements on a device with \SI{10}{\micro\meter} $\times$ \SI{12}{\micro\meter} active area.

Optical excitation is performed with a \SI{785}{\nano\meter} laser (Toptica, iBeam smart), which is focused onto the SiC device using an NA=0.65 objective (Zeiss, Plan-Achromat 40$\times$).
The sample is mounted on a 3D piezo stage with \SI{100}{\micro\meter} travel range (Physik Instrumente, P-561.3CD).
A 3D Helmholtz coil arrangement is used for applying magnetic fields in arbitrary directions.
Radiofrequency (RF) for spin control and manipulation are provided by a rubidium-referenced (EFRATOM, LPRO-101) signal generator (Rohde \& Schwarz, SMIQ03B), pulsed by a microwave switch (Mini Circuits, ZASWA-2-50D), amplified (Mini Circuits, ZHL4240W) and finally applied via a coplanar waveguide on the printed circuit-board sample holder below the sample.
This sample holder also incorporates contact pads, to which the device contacts are wire-bonded.
For better SNR, we use a lock-in detection scheme (Stanford Research, SR830).
Therefore the signal is locked to the laser pulses for photocurrent measurements and on the modulated RF pulses for ODMR, PDMR and Rabi measurements).
As the RF pulses are short (\SI{300}{\nano\second}), the locking is achieved by repeating the whole spin control pulse sequences with and without RF multiple times at a lock-in frequency of \SI{429}{\hertz}, as depicted in Fig.~\ref{fig:pulses}(d).
Typical pulse lengths for optical initialization in PDMR are \SI{600}{\nano\second} laser pulse followed by \SI{1}{\micro\second} settling time.

To measure a spin-dependent photocurrent, a bias voltage is applied using a source measure unit (Keithley, 2636B).
The resulting photocurrent is converted to a voltage by a  transimpedance amplifier (Femto, DLPCA-200, gain of $10^8$ for PDMR, $10^9$ for Rabi), which is low-pass filtered at \SI{1}{\kilo\hertz}.
By scanning the sample position, we record photocurrent maps.
At each position, we measure the photocurrent as a function of excitation power and fit the recorded data with a second order polynomial function to infer the contributions of single- (linear) and two-photon (quadratic) processes. 
For ODMR measurements, we detect fluorescence emission from \SI{850}{\nano\meter} to \SI{950}{\nano\meter} using a photodiode (Newport, Model 2151) and feed the signal directly into the lock-in amplifier.
All measurements are performed at a laser power of \SI{178.5}{\milli\watt} (unless  stated otherwise) in order to keep the same experimental conditions for PDMR and ODMR.
The beam in front of the photodiode is attenuated by an iris to prevent detector saturation.
For PDMR measurements, the output of the transimpedance amplifier is connected to the lock-in amplifier instead of the photodiode.
To avoid artefacts due to frequency-dependent coupling into the SiC device, we keep the RF frequency constant and stepwise change the magnetic field \BO{} revealing the magnetic resonance induced signals.
The magnetic field is roughly aligned along the $c$-axis of the sample.
In order to map the PDMR signal, we repeatedly measure and average the PDMR amplitude.
This is done by subtracting the off-resonant signal from the on-resonance data.
The off-resonant signal is obtained at a \BO -field strength corresponding to \SI{23}{\mega\hertz} detuning.

A similar approach is used for spin Rabi oscillation measurements. Here,  a fixed \BO{} field is applied and a RF field ($B_1$ field) at the spin resonance frequency drives the system, while the RF pulse length is altered and the overall sequence duration is kept constant.
To account for potential RF pick-up by the lock-in scheme, we subtract an off-resonant baseline signal as described for the PDMR mapping.

Fig.~\ref{fig:sample}(b) shows the photocurrent map of our device.
We find that the response is localized inside the center of the device.
The spatial map of the contribution of two-photon process in the photocurrent extracted from quadratic fitting of the laser power dependency of photocurrent data (see Fig.~\ref{fig:sample}(c)) is shown in Fig.~\ref{fig:sample}(d).
Comparing Fig.~\ref{fig:sample}(b) with Fig.~\ref{fig:sample}(d), we find that most areas show mainly linear response, indicating single photon absorption from shallow traps.
In the center of the device, we observe a pronounced quadratic dependence.
We perform all further measurements in this area.

\begin{figure}
\includegraphics[width=1\linewidth]{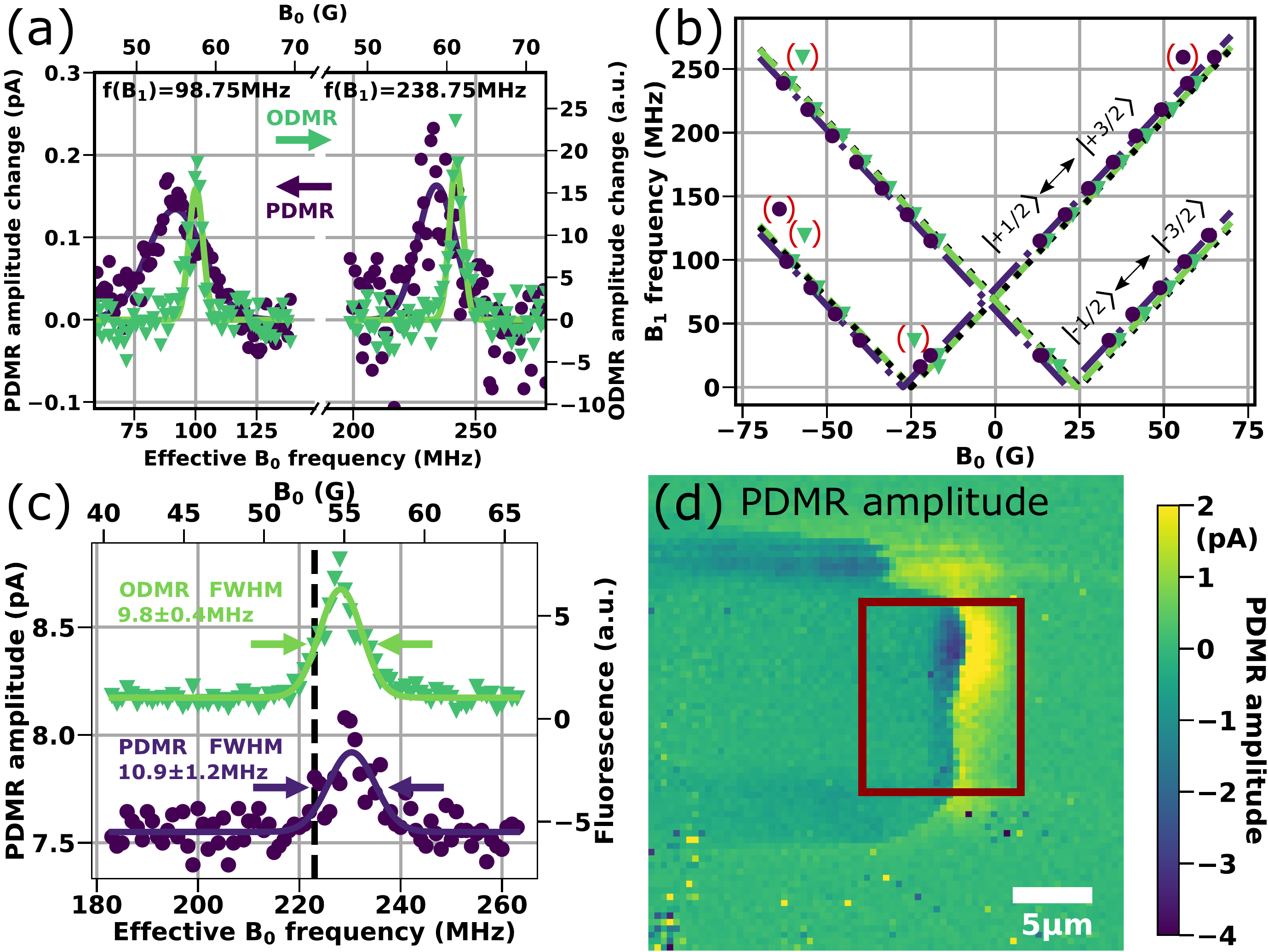}
\caption{\label{fig:pdmr}ODMR vs. PDMR (a) Comparison of \VSi{} PDMR and ODMR signals of upper and lower transition between \pmSOH{} and \pmSTH{} spin subsets at -\SI{10}{\volt} Bias. Offset removed for comparison. Green triangles indicate ODMR, purple dots PDMR data points. Gaussian fits to PDMR are plotted as purple, ODMR as green lines, respectively. (b) PDMR and ODMR Zeeman splitting. Dashed colored lines show a fit to the data, where data points in brackets are neglected. Black colored dotted lines show expected theoretical values and are overlaping with measurement results. (c) Similar linewidths in ODMR and PDMR measurements at +\SI{20}{\volt} Bias. The black dashed line indicates the expected resonance frequency. (d) PDMR amplitude map over optical excitation position. Estimated device position marked as rectangle.}
\end{figure}

We subsequently perform stepwise \BO -field dependent measurements at fixed RF frequencies of \SI{98.75}{\mega\hertz} and \SI{238.75}{\mega\hertz} to resolve the  resonances of \mSOH{} $\leftrightarrow$ \mSTH{} and \pSOH{} $\leftrightarrow$ \pSTH{} transitions, respectively.
Both PDMR and ODMR results are shown in Fig.~\ref{fig:pdmr}(a), exhibiting the expected magnetic resonance for both ODMR and PDMR except for a small difference in the resonance positions.
We tend to attribute the shift between ODMR and PDMR resonance to offset fields present in the device, probably due to the proximity to the ferromagnetic Ni contacts (see Supporting Information).
As an ensemble is used and field inhomogenities are present, lines are expected to be of Gaussian shape envelope.
The linewidths in Fig.~\ref{fig:pdmr}(a) are much broader in the electrical case compared to the optically detected lines.
We attribute this to a mismatch in the detection volumes for both techniques  in combination with the residual magnetization of the Ni contacts.
As the fluorescence light is not spatially filtered in case of the electrical readout, the detected signal may differ in position compared to the ODMR measurements.
Nonetheless, data recorded at a different position shown in Fig.~\ref{fig:pdmr}(c) suggests similar linewidths for PDMR and ODMR and thus proves that broadening in Fig.~\ref{fig:pdmr}(a) is not due to the PDMR technique (see Supporting Information).
To check that the measured signal originates from V$_2$ centers, we measure the ground state ZFS via Zeeman splitting measurements by observing the resonances at various magnetic fields.
As depicted in Fig.~\ref{fig:pdmr}(b), the ZFS is found by fitting the model function $f_{\textnormal{res}}(B_0)=\left|\textnormal{ZFS}\pm g \textnormal{µ}_\textnormal{B} \left(B_0+B_\textnormal{offset}\right)\right|$ to the magnetic field dependence of the resonances.
Thereby,  $g$ is the effective electron Landé factor for \VSi{}, µ$_\textnormal{B}$ is the Bohr magneton and $B_\textnormal{offset}$ is a magnetic field accounting for the device internal fields.
With this, we find $\textnormal{ZFS}_{\textnormal{ODMR}}=$~\sipm{69.0}{0.3}{\mega \hertz} and $\textnormal{ZFS}_{\textnormal{PDMR}}=$~\sipm{69.1}{0.3}{\mega\hertz} for the optically and electrically detected case, respectively.
The $g$ factors found by the fit are $g_{\textnormal{ODMR}}=$~\sipm{2.02}{0.01}{} for ODMR and $g_{\textnormal{PDMR}}=$~\sipm{2.03}{0.01}{} for PDMR, while the offset field is $B_\textnormal{offset,ODMR}=$~\sipm{0.4}{0.1}{} and $B_\textnormal{offset,PDMR}=$~\sipm{3.0}{0.1}{\gauss}, respectively.
The data presented corroborate that the signal originates from V$_2$ centers.
As shown in Fig.~\ref{fig:pdmr}(d), the PDMR signal is located in the same area where the two-photon photocurrent contribution was found in Fig.~\ref{fig:sample}(d).
\begin{figure}
\includegraphics[width=1\linewidth]{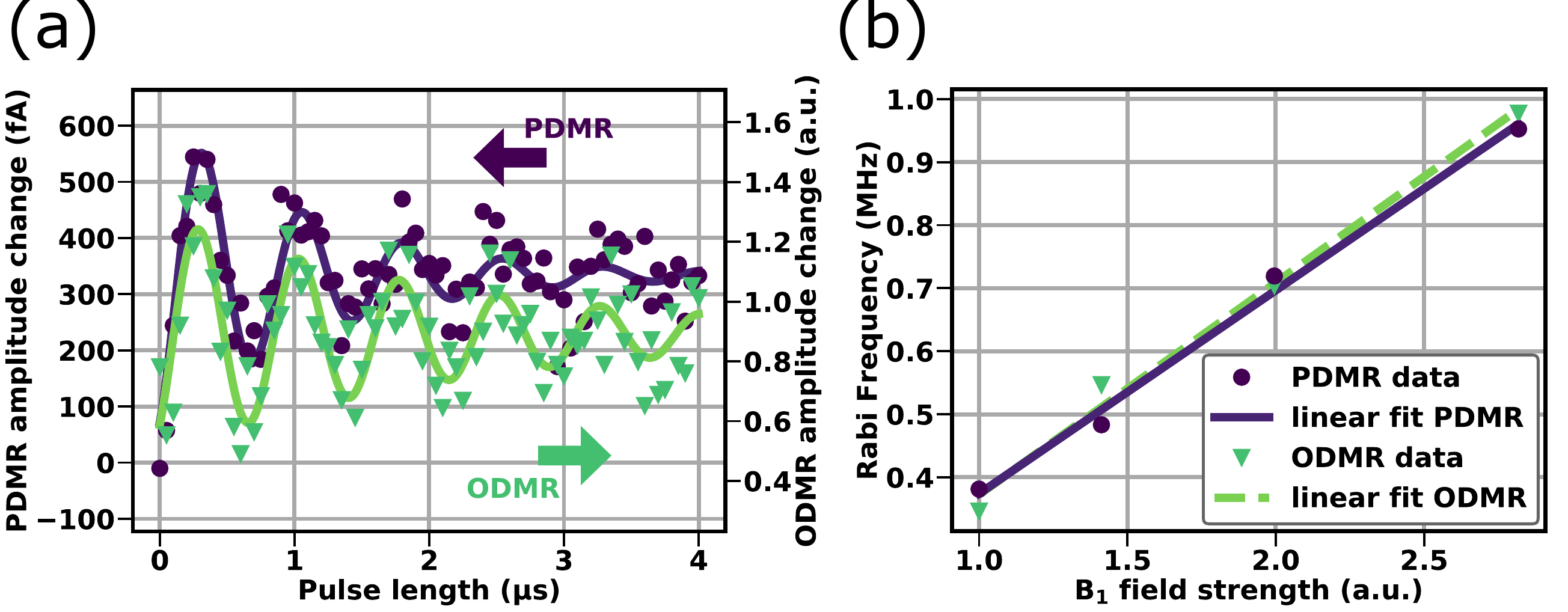}
\caption{\label{fig:rabi} 
(a) Electrically detected Rabi oscillation directly compared to optically detected Rabi oscillation. (b) Driving field strength dependence of Rabi oscillation frequency for optical and electrical readout.}
\end{figure}
However, we find the sign of the PDMR signal to be dependent on the location within the device  as can be seen in Fig.~\ref{fig:pdmr}(d).
We tentatively attribute this to a change in the Fermi level in the device caused by 
charge state and ionization processes of surrounding defects~\cite{magnusson_excitation_2018}.
As a result, we cannot clearly determine if excitation from the ES or MS is responsible for the observed PDMR effect in the present device.

Next, we demonstrate coherent control, which is at the heart of advanced quantum control protocols.
To this end, we first initialize the GS spin population into the \pmSOH{} subspace via optical excitation.
Subsequently a RF driving pulse of variable length to the \pSOH{} $\leftrightarrow$ \pSTH{} transition is applied. Finally the spin state is read out either optically or electrically using the next laser pulse.
The latter at the same time ensures that the system is re-initialized for the following  cycle.
Experimental results for both ODMR and PDMR recorded under identical measurement conditions are shown in Fig.~\ref{fig:rabi}(a). 
We observe Rabi oscillations with essentially identical oscillation frequency and same-order decay times from both detection methods, which indicates that PDMR has no major detrimental effect on dephasing of the continuously driven system.
We further record the Rabi oscillation frequency as a function of RF field strength and observe the expected linear increase (see Fig.~\ref{fig:rabi}(b)).
This proves that the PDMR of the \VSi{} spin state in SiC allows for coherent spin manipulation and readout of the ground state and thus fulfills the fundamental requirements for more complex quantum control schemes.

To evaluate the performance of the PDMR technique, we performed a parameter dependency study (see Supporting Information).
We find a ten-fold increase in SNR in ODMR compared to PDMR after normalizing to the same measurement time. 
In addition, the PDMR contrast is around one order of magnitude smaller than the ODMR contrast with the current device. While PDMR amplitudes are in the range of \si{\pico\ampere}, the mean dc background current measured by an oscilloscope parallel to the lock-in amplifier is on the order of a few \si{\nano\ampere}.
This results in a typical contrast of \SI{0.03}{\percent}.
On the other hand, ODMR measurements yield a contrast of around \SI{0.1}{\percent}.
The background current mainly consists of  the resistive current through the device due to the bias.
The laser induced photocurrent also contributes to the background, but due to the pulsed type of measurement is decreased by the duty cycle.
However, our measurements suggest that we are limited by the current experimental conditions and that multiple parameters can still be optimized (see Supporting Information).
Especially with increasing laser power the ODMR contrast saturates, whereas no saturation behavior is observed for PDMR yet.
This is consistent with findings for NV ensembles in diamond~\cite{bourgeois_photoelectric_2015}.
Furthermore, refining the measurement technique and device structure can potentially improve SNR.
A large contribution to the noise floor is stray RF fields.
We anticipate a gain in SNR by improving the device structure to be more resilient against parasitic RF coupling.
In addition, the stepwise measurement was done in a conservative way and seconds of settling time between magnetic field steps were chosen in order to reach a quasi-static situation, while lock-in integration time was set to \SI{30}{\milli\second}.
Using a real magnetic-field sweep or frequency-modulated RF field will speed up signal accumulation.
However, due to the RF-frequency-dependent stray currents and no possibility to directly sweep the magnetic field in our experimental conditions, we have not incorporated such techniques yet.
Moreover, changes to the doping profile may allow to enhance carrier extraction efficiency, but may come with the cost of an increase in background photocurrent.
As the large bandgap hinders a two-photon band-to-band excitation with a \SI{785}{\nano\meter} laser, the background photocurrent is likely generated by excitation of other intra-band defects created besides the \VSi{} ensemble during the electron irradiation.
As the background limits transimpedance gain, a trade-off between signal extraction efficiency and background has to be found.
Another parameter is device geometry, e.g. channel width and thickness of the active layer.
By this, the detection volume might be enlarged and leakage currents further reduced.
Interestingly, only a small area within the aperture shows contribution to PDMR, although the details of the process have to be understood first.
To this end, we suggest to measure the dependence of the signal on excitation laser wavelength and pulse length, which might give insight into the ionization process and may ultimately improve readout fidelity and state preparation~\cite{aslam_photo-induced_2013,siyushev_optically_2013}.
Since we have shown that coherent spin control of \VSi{} can be combined with PDMR, phase interferometry type sensing protocols can be utilized, which can boost sensitivity in metrology applications by many orders of magnitude
\cite{schirhagl_nitrogen-vacancy_2014,taylor_high-sensitivity_2008}.

In summary, we have demonstrated photo-electrical readout of a \VSi{} spin ensemble in a 4H-SiC metal-semiconductor-metal device under ambient conditions. 
We also report electrically detected spin coherence of this ensemble.
This underlines the great potential of SiC and PDMR for quantum applications.
The availability  of large wafer production and processing techniques are very promising to future integration of electrical quantum devices at an industrially relevant scale.
Advanced fabrication techniques can be used to integrate \latin{e.g.} high-performance CMOS transimpedance amplifiers on-chip~\cite{djekic_0.1_2018}. This would allow miniaturization and quantum device integration into a classical circuit design.
Even integration of the optical light source might be feasible in the future~\cite{lu_white_2017}.
Altogether, this work provides a first step towards integrated electrical quantum devices in 4H-SiC for quantum technology.

\section{Author contributions}
The initial planning of the project was done by M.N., M.W., S.-Y.L, N.T.S. and J.W..
J.U-H. and N.T.S. designed the device structure and performed the sample growth.
R.S. fabricated the device. 
S.O., T.O. and J.I. planned and performed electron irradiation.
M.N. designed and performed all experiments.
M.N., T.R., M.W., A.M., N.M., Y-C.C., S.-Y.L and J.W. analyzed the data.
The manuscript was written by M.N., M.W., T.R., R.S. and J.W. with contribution from all of the authors. 

\section{Acknowledgments}
The authors thank Florian Kaiser for his valuable help in preparing the manuscript.
We acknowledge financial support by the EU (ASTERIQS and ERC SMeL), BMBF (BrainQSens), the Max Planck Society, the Volkswagen Foundation, the Swedish Research Council (VR 2016-04068 and VR 2016-05362), the Carl Tryggers Stiftelse f\"or Vetenskaplig Forskning (CTS 15:339), the Swedish Energy Agency (43611-1), the Knut and Alice Wallenberg Foundation (KAW 2018.0071), the Korea Institute of Science and Technology institutional programs (2E27231, 2E29580) and the Japan Society for the Promotion of Science KAKENHI (17H01056).
\section{Abbreviations}
CB -- conduction band \newline
ES -- excited state \newline
GS -- ground state \newline
ISC -- intersystem crossing \newline
MS -- metastable state \newline
MSM -- metal-semiconductor-metal structure \newline
NV -- nitrogen-vacancy center in diamond \newline
ODMR -- optically detected magnetic resonance \newline
PDMR -- photocurrent detected magnetic resonance \newline
RF -- radiofrequency \newline
SiC -- silicon carbide  \newline
SNR -- signal-to-noise ratio \newline
TIA -- transimpedance amplifier \newline
VB -- valence band \newline
\VSi{}, V$_\textnormal{2}$ -- negatively charged silicon-vacancy (at cubic lattice site) \newline
ZFS -- zero field splittings

%\section{Notes}
%The authors declare no competing financial interest.
%\begin{suppinfo}
%Experimental data for conductivity before and after irradiation, PDMR maps at different depth %and crossection, PDMR data dependent on bias voltage, RF and laser power, discussion on signal %contrast and SNR, calculation of dc magnetic field sensitivity, ODMR detected resonance %frequencies along $z$-axis and $x$-axis
%\end{suppinfo}

%\bibliography{pdmr_paper_arxiv}

%\documentclass[manuscript=suppinfo]{achemso}
\setkeys{acs}{articletitle = true}

%\usepackage{float}
%\usepackage{amsmath} 
%\usepackage{xcolor}
%\usepackage{hyperref}
%\usepackage{graphicx}
%\usepackage{siunitx}
%\hypersetup{colorlinks,linkcolor={blue},citecolor={blue},urlcolor={red}}  
%\newcommand{\citeColored}[2]{{\hypersetup{citecolor=#1}#2}}

%\newcommand{\comment}[1]{{\textcolor{red}{\textbf{[#1]}}}} 
%\newcommand{\valuemissing}[1]{{\textbf{#1}}} 
%\newcommand{\refmissing}[1]{\textcolor{red}{\citeColored{red}{#1}}}
%\newcommand{\refcheck}[1]{\textcolor{orange}{\citeColored{orange}{#1}}}
%\newcommand{\refweak}[1]{\textcolor{blue}{\citeColored{blue}{#1}}}

%\newcommand{\VSi}{V$_\textnormal{Si}^{-}$}
%\newcommand{\STH}{3/2}
%\newcommand{\SOH}{1/2}

%\newcommand{\BO}{$\textnormal{B}_\textnormal{0}$}

%\newcommand{\pmSTH}{$\pm$3/2}
%\newcommand{\pmSOH}{$\pm$1/2}

%\newcommand{\pSTH}{$+$3/2}
%\newcommand{\pSOH}{$+$1/2}
%\newcommand{\mSTH}{$-$3/2}
%\newcommand{\mSOH}{$-$1/2}

%\newcommand{\sipm}[3]{#1 $\pm$ \SI{#2}{#3}}

%\DeclareSIUnit\gauss{G}
%\DeclareSIUnit{\belmilliwatt}{Bm}
%\DeclareSIUnit{\dBm}{\deci\belmilliwatt}

\title{Supporting Information}
%\title{Coherent electrical readout of defect spins in 4H-SiC  by photo-ionization at ambient conditions}
%\author{Matthias Niethammer}
%\email{matthias.niethammer@pi3.uni-stuttgart.de}
%\author{Matthias Widmann}
%\author{Torsten Rendler}
%\author{Naoya Morioka}
%\author{Yu-Chen Chen}
%\author{Rainer St\"ohr}
%\affiliation{3rd Institute of Physics and Center for Applied Quantum Technologies, University of Stuttgart, 70569 Stuttgart, Germany}
%\author{Jawad Ul Hassan}
%\affiliation{Department of Physic, Chemistry and Biology, Link\"oping University, SE-581 83 Link\"oping, Sweden}

%\author{Shinobu Onoda}
%\author{Takeshi Ohshima}
%\affiliation{National Institutes for Quantum and Radiological Science and Technology, Takasaki 370-1292, Japan}

%\author{Sang-Yun Lee}
%\affiliation{Center for Quantum Information, Korea Institute of Science and Technology, Seoul 02792, Republic of Korea}
%\author{Amlan Mukherjee}
%\affiliation{3rd Institute of Physics and Center for Applied Quantum Technologies, University of Stuttgart, 70569 Stuttgart, Germany}
%\author{Junichi Isoya}
%\affiliation{Faculty of Pure and Applied Sciences, University of Tsukuba, Tsukuba 305-8573, Japan}
%\author{Nguyen Tien Son}
%\affiliation{Department of Physic, Chemistry and Biology, Link\"oping University, SE-581 83 Link\"oping, Sweden}
%\author{J\"org Wrachtrup}
%\affiliation{3rd Institute of Physics and Center for Applied Quantum Technologies, University of Stuttgart, 70569 Stuttgart, Germany}
%\alsoaffiliation{Max Planck Institute for Solid State Research, 70569 Stuttgart, Germany}

\date{\today}
%\begin{document}
\def\theequation{S\arabic{equation}}
\def\thefigure{S\arabic{figure}}
\setcounter{equation}{0}
\setcounter{figure}{0}
%\maketitle

\newpage
\section{Supporting information}

\subsection{Sample growth and fabrication details}
Different layers were grown by chemical vapor deposition (CVD) on \SI{4}{\degree} off-axis n-type 4H-SiC substrate. The first layer grown is a \SI{10}{\micro\meter} thick semi-insulating V-doped layer followed by a n$^{-}$ layer (\SI{10}{\micro\meter}) with a free carrier concentration  of \SI[mode=math]{1e14}{\centi\meter ^{-3}} at room temperature. The top n$^{++}$-type N-doped contact layer is \SI{400}{\nano\meter} thick with a doping concentration of  \SI[mode=math]{8e17}{\centi\meter ^{-3}}.

\begin{figure}[H]
\centering
\includegraphics[width=\linewidth]{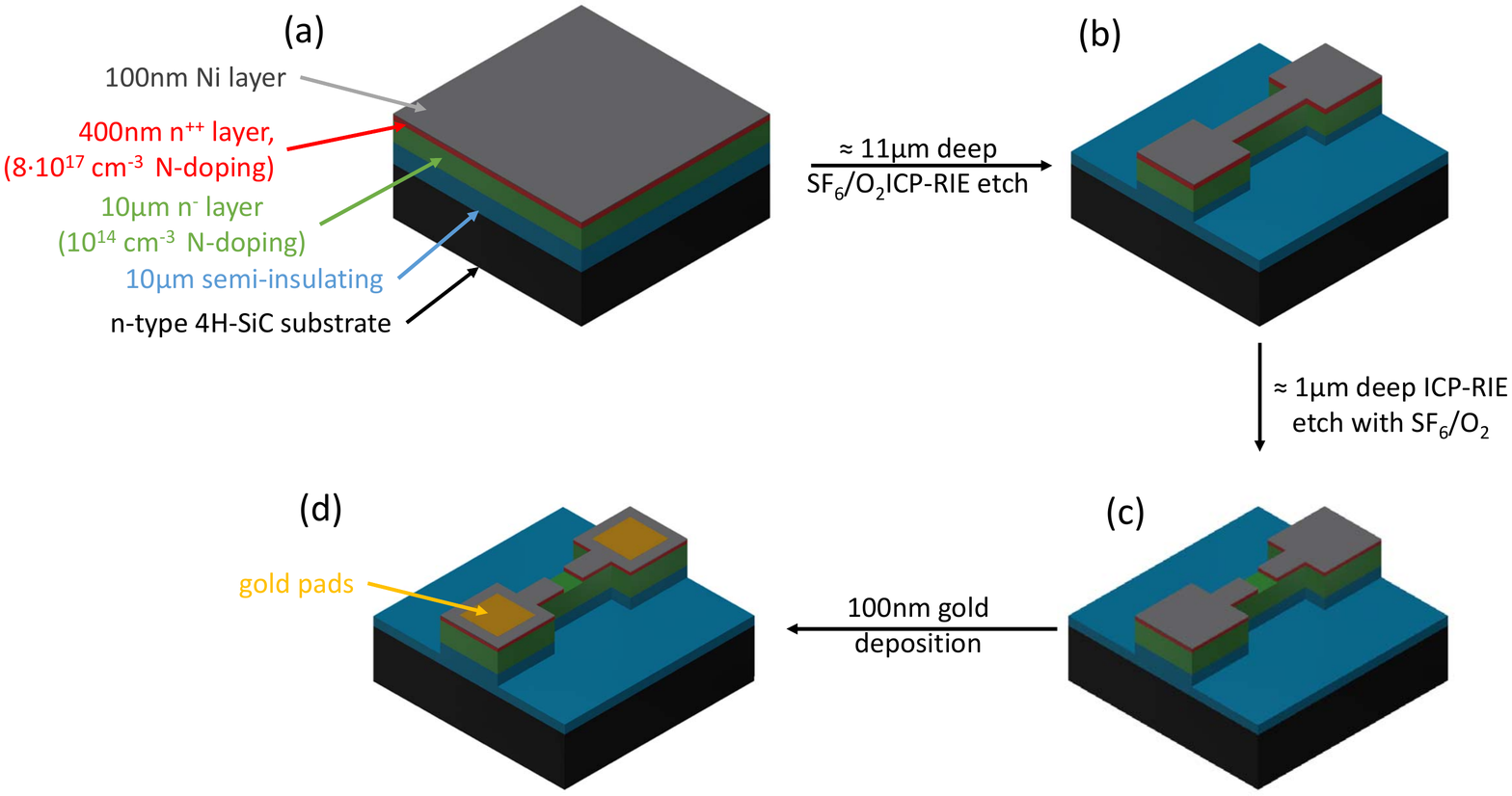}
\caption{\label{fig:fabric} Device structure and schematic representation of the device fabrication process. (a) Initial sample layout. (b) Device geometry after the first deep etch step, isolating the devices from others. (c) Removal of the n$^{++}$ layer between the contacts to allow optical access. (d) Final device configuration with two gold pads for wirebonding.}
\end{figure}

Fig.~\ref{fig:fabric}(a) shows the sample layout before structuring. While the Ni layer serves as a Schottky contact in the final device, it is also used as an etching mask during the fabrication process. All dry etching steps are performed using an ICP-RIE with SF$_6$/O$_2$ gas mixture, which realizes high etching selectivity of SiC over Ni. In the first step, all Ni outside the device region is removed. The sample is then plasma-etched $\approx$\SI{11}{\micro\meter} deep, which removes the n$^{++}$ layer and the n$^{-}$ layer and stops in the semi-insulating layer (Fig.~\ref{fig:fabric}(b)). During this step, the device region is protected by the residual Ni layer. Subsequently, Ni is removed in a rectangular region between the two contact pads of the device. This allows to etch the n$^{++}$ layer in this region by a short SF$_6$/O$_2$ ICP-RIE plasma step. The etching depth is chosen to be roughly twice the thickness of the n$^{++}$ layer, ensuring that it is completely removed. Finally, gold pads are deposited on both contacts for wirebonding, resulting in the final device as depicted in Fig.~\ref{fig:fabric}(d).

\subsection{$I$--$V$ Characteristics of the device used}
The $I$--$V$ characteristics of the device have been measured before and after electron irradiation.
Note that these were measured on different experimental setups. For the measurement before irradiation, a source measure unit (SMU, Keithley, 487) with a manual probe station was used.
Measurements after irradiation were performed with the PDMR setup using a SMU (Keithley, 2636B). The sample was mounted on a PCB sample holder. Connections between sample and PCB were wirebonded.

\begin{figure}[H]
\centering
\includegraphics[width=1\linewidth]{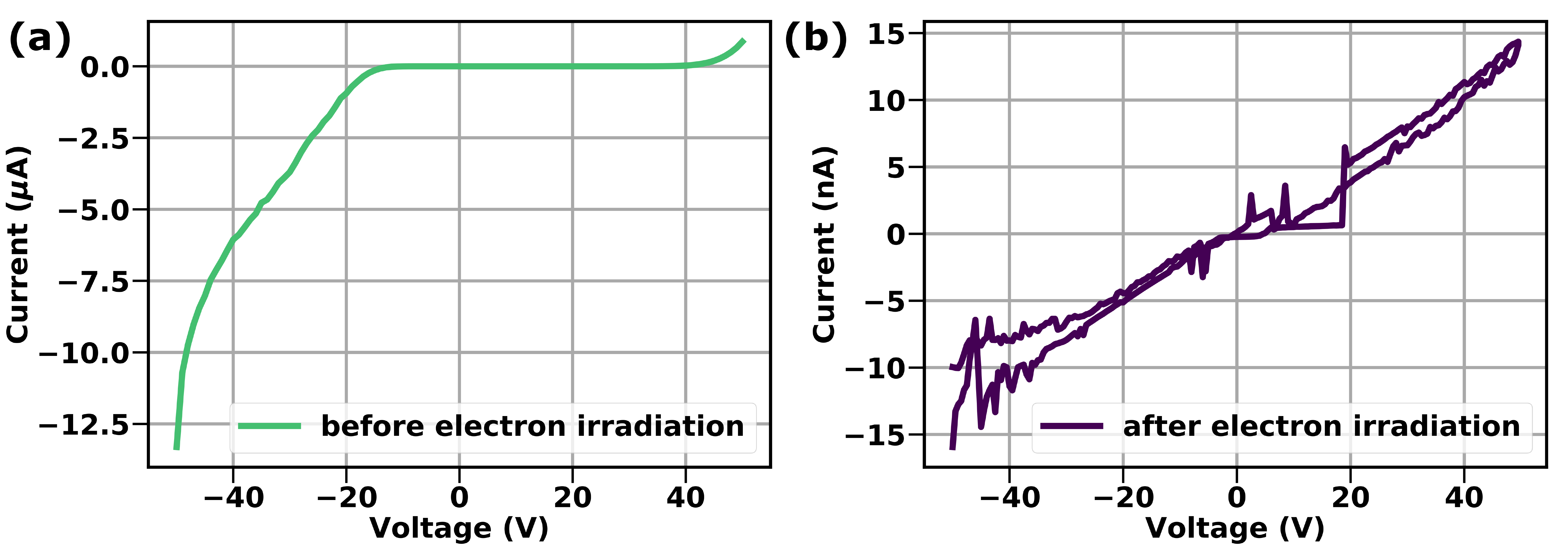}
\caption{\label{fig:irrad} $I$--$V$ characteristics of the device. (a) Before electron irradiation. (b) After electron irradiation.}
\end{figure}

As shown in Fig.~\ref{fig:irrad}(a), the device shows rectifying behavior at positive and negative bias conditions. Thus, the contacts are assumed to be Schottky type.
Then the sample has been irradiated by electrons with a dose of \SI[mode=math]{1e17} {\centi\meter ^{-2}} and an energy of \SI{2}{\mega\electronvolt}.
The measurement of the irradiated device clearly shows over 2 orders of magnitude less conductivity compared to the non-irradiated device (see Fig.~\ref{fig:irrad}(b)), which we attribute to the doping compensation due to irradiation-induced defects~\cite{ohyama_radiation_2005}.

\subsection{Position mapping of PDMR signal}
\begin{figure}[H]
\centering
\includegraphics[width=0.75\linewidth]{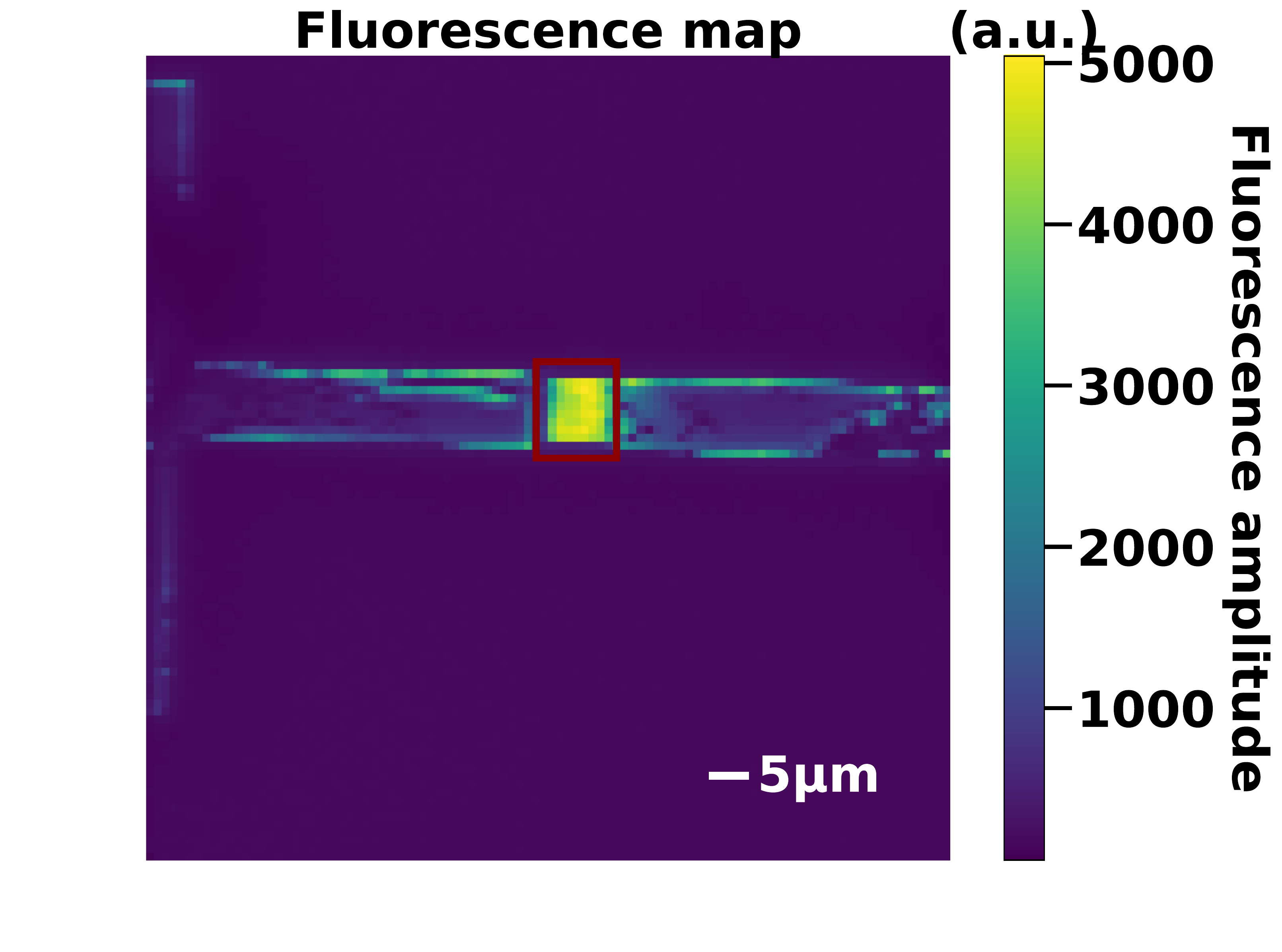}
\caption{\label{fig:refl_map} Fluorescence detected $xy$-scan of the device. The rectangular aperture is clearly visible. By this we identify the measurement positions shown in main text in Fig.~1(b,d) and Fig.~3(d).}
\end{figure}

To map the position of the obtained PDMR signal to the device structure, we utilize laser scanning. Thereby we subsequently acquire PDMR signal and fluorescence emission of the \VSi{} ensemble.
The fluorescence locates the device within the given scan range of \SI{100}{\micro\meter} as shown in Fig.~\ref{fig:refl_map}.
Because the optical detection is performed without spatial filtering, a slight offset in position may exist.

\begin{figure}[H]
\centering
\includegraphics[width=1\linewidth]{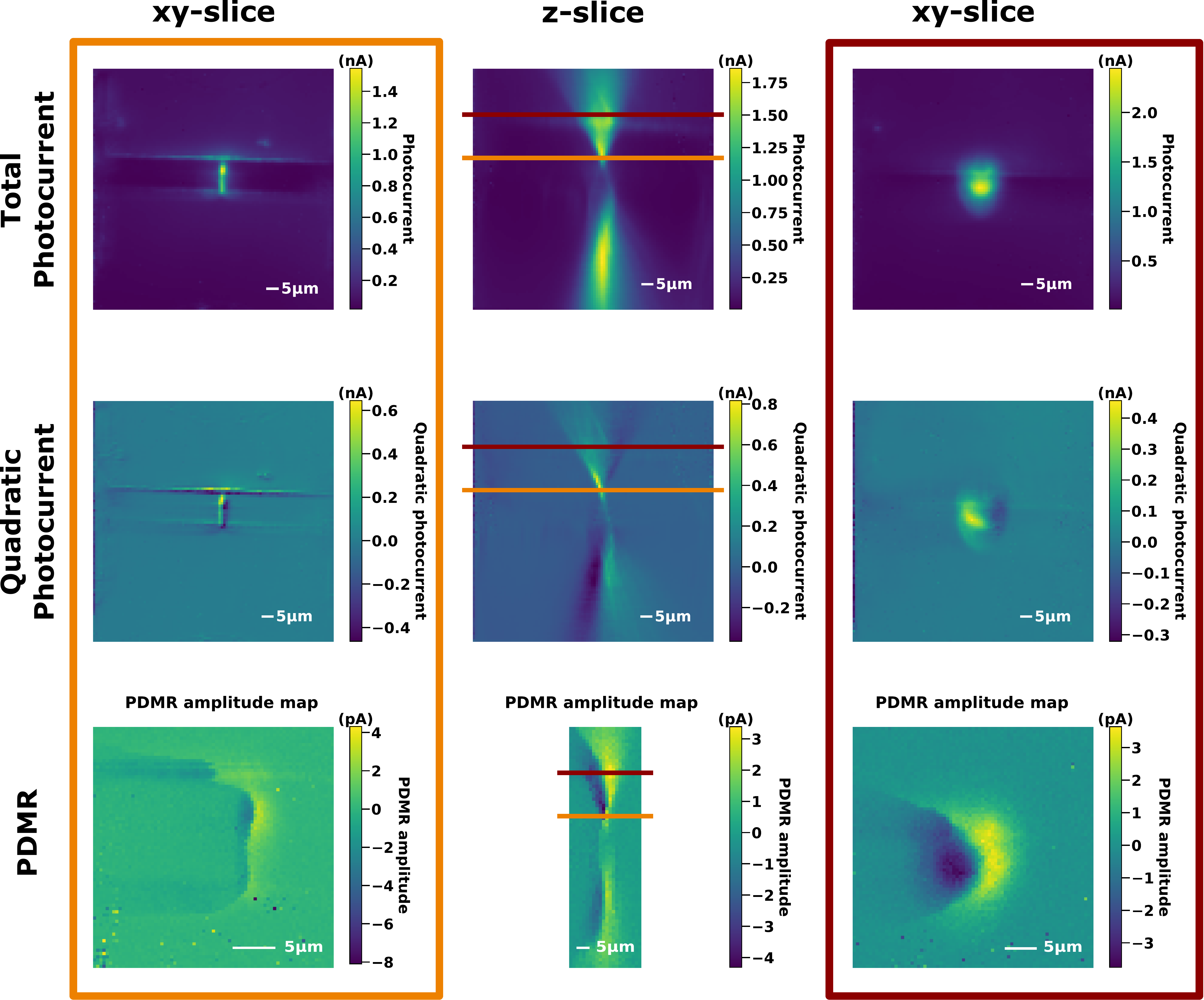}
\caption{\label{fig:pdmr_maps} Photocurrent, two-photon contribution to photocurrent and PDMR amplitude maps recorded at $+$\SI{20}{\volt} bias. Orange and red lines in center z-slices indicate focal position for on-focus (orange) and out-of-focus (red) recorded maps.}
\end{figure}

All measurements in the main text have been performed at a fixed depth for consistency.
Crossections of photocurrent and PDMR signals are given in Fig.~\ref{fig:pdmr_maps}.
The $z$-slice shows that both photocurrent and PDMR amplitude are dependent on the focal position.
The $xy$-slices show a thin strip (marked orange) of effective photocurrent generation that evolves to a larger area when defocusing (marked red).
When the focus is inside the device, we do not find a PDMR signal.
We attribute this finding to a small excitation volume, which results in a too small number of defects involved in the PDMR process.
As increasing the excitation area, we pick up a measureable PDMR signal.
However, due to the decrease in laser power density, the signal does not saturate.
At the moment, it is unclear to us why this process only appears at the center of the device.
A convolution of excitation volume and active area should be the expected result.

\subsection{Parameter dependencies: Bias voltage, laser power and RF power}
\begin{figure}[H]
\centering
\includegraphics[width=1\linewidth]{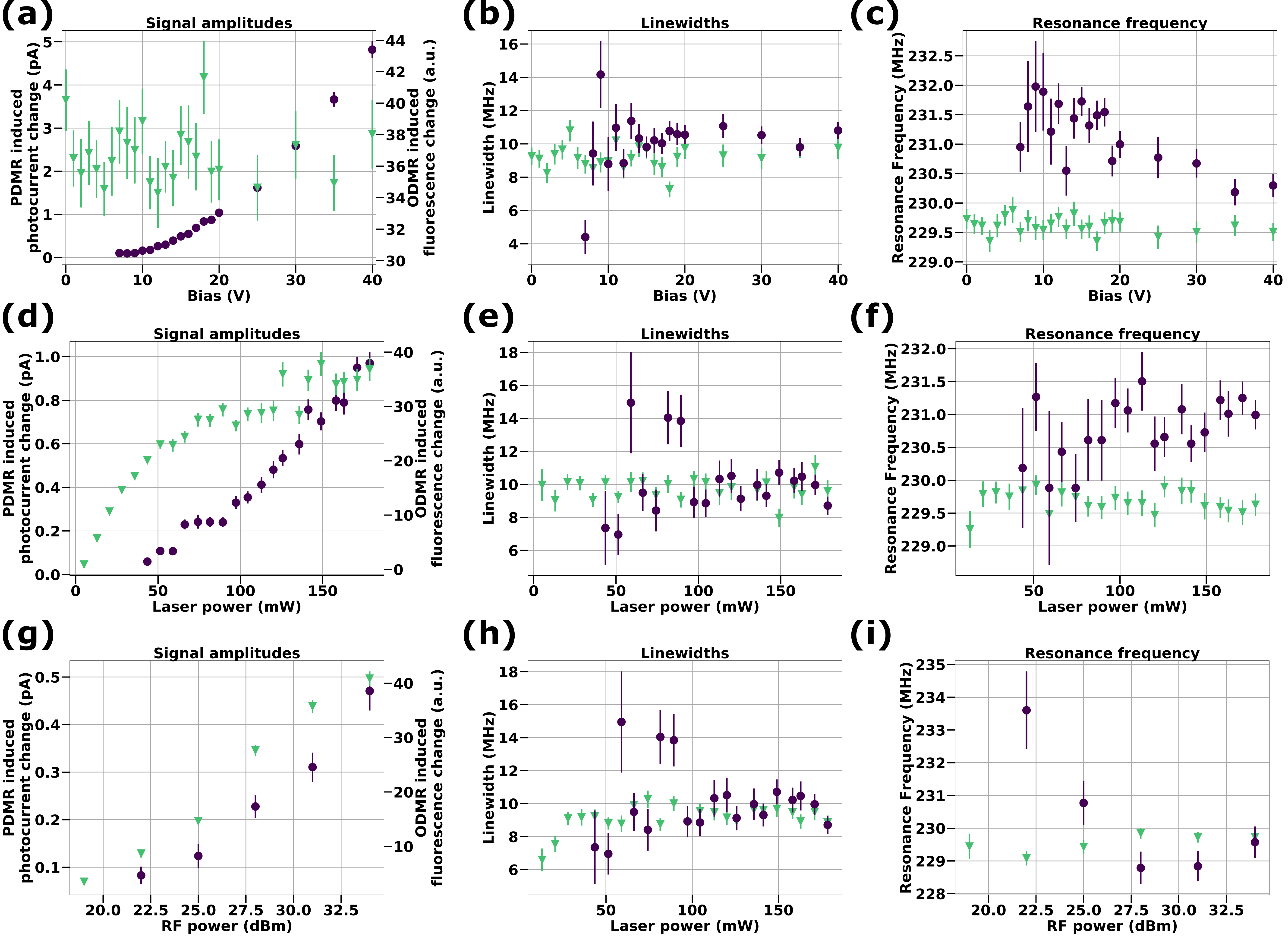}
\caption{\label{fig:parameters} Parameter dependence comparison for ODMR (green) and PDMR (purple).
Data points are obtained from Gaussian peak fit. Meausurement conditions: Bias $+$\SI{20}{\volt}, \SI{178.5}{\milli\watt} laser and \SI{34}{\dBm} RF power, \SI{223}{\mega\hertz} RF frequency.}
\end{figure}

Fitting a single resonance peak in a PDMR measurement to the $m_s=$\pSOH{}$\leftrightarrow$\pSTH{} transition gives information on the signal amplitude, linewidth, resonance frequency and the offset signal.
The offset signal is mainly caused by stray RF field.
Fig.~\ref{fig:parameters} compares the dependency of ODMR and PDMR signal, on bases of amplitude, linewidth and resonance frequency when bias, laser power and RF power are changed.
Note that although measurements have been performed under the same conditions for ODMR and PDMR, a slight offset in detected volume is possible (see discussion in main text).
PDMR signals at small laser or RF powers as well as small bias voltages show low SNR and thus larger errorbars.
Errorbars are obtained from standard error of least-square fitting, neglecting the noise contribution at each measurement point. 

The bias voltage does not affect the ODMR signals.
On the other hand, the PDMR amplitude increases monotonically without saturating.
The applied bias determines the charge extraction efficiency, and larger bias is expected to increase the measured signal amplitude. 
Note that linewidth for ODMR and PDMR are similar.
A small shift of resonance frequency dependent on bias is visible for PDMR (see Fig.~\ref{fig:parameters}(c)), but not in ODMR. This shift appears in the low bias regime in which the SNR is poor.
At this moment, we do not have an explanation for this behavior.

The incident laser power does change neither linewidth nor resonance frequency.
However, the fitted peak amplitude increases for both ODMR and PDMR.
This observation is consistent with the suggested mechanism for the PDMR for the \VSi{} in SiC because stronger optical excitation will enhance the photo-ionization probability (see Fig.~1(a)-(c) of the main text).
Note that laser power dependence suggests a saturation behaviour for ODMR, while no saturation could be achieved in the PDMR case.

Larger RF power increases linewidth of the magnetic resonance due to power broadening.
In case of PDMR, a larger shift in resonance frequency is only visible for the point at smallest RF power at which signal is still picked up in PDMR.
At this data point, the signal has very low amplitude.
The amplitudes of the PDMR and ODMR signals increase with larger RF power.
Assuming inhomogeneous broadening, increasing the RF power increases the excitation bandwidth and thus more defects contribute to the signal.
No saturation of the microwave transition is visible. Note that the point of highest applied RF power is already measured in the RF amplifier compression regime.

\subsection{Discussion on PDMR contrast, SNR and sensitivity}
In a typical ODMR experiment, the contrast $c$ is defined as the spin-dependent fluorescence change $\Delta_{\textrm{PL}}$ at resonance to the off resonance fluorescent signal PL$_{\textrm{BG}}$ (baseline):
\begin{equation}
\label{eqn:contrast_odmr}
	c= \Delta_{\textrm{PL}}/\textrm{PL}_{\textrm{BG}}.
\end{equation}
As this definition fits the requirement, as long as an absolute signal is acquired, this is not directly applicable in case of lock-in detection, as only a change in an acquired quantity is detected. In other words, the absolute measure of the given input is lost, which is essential for the former definition of contrast, and only changes modulated by the lock-in frequency are detected. 
Nevertheless, the detected lock-in signal contains a constant offset.
Here, as the device is in close proximity to the co-planar waveguide, the offset is dominated by a frequency-dependent coupling of the RF field to the device, which is modulated exactly at the lock-in frequency. However, the use of this offset in the definition of contrast as the baseline would lead to a non-physical interpretation of PDMR contrast.
Hence, one could give a device-specific contrast only, which compares the amount of RF coupling for a given RF frequency with PDMR signal.
Based on the original definition of contrast, we further extend this definition by comparing the amplitude to the maximum acquired signal as follows:
\begin{equation}
\label{eqn:contrast_pdmr}
 c = \textrm{A} / \textrm{max}( \left|\textrm{BG}\right|, \left|\textrm{BG}+\textrm{A}\right|). 
\end{equation}
Here, BG is the fluorescence background or dc offset of the PDMR signal and A is the ODMR or PDMR amplitudes.
By this definition, the maximum achievable contrast by fluorescence is limited to 100\%, resulting in a more meaningful quantity. 
We simultaneously monitor the detected PDMR signal by lock-in detection and use an oscilloscope in parallel to the lock-in amplifier in order to detect the mean magnitude of the dc signal.
This dc offset additionally to the previous signals is composed of bias and photocurrent contribution. The lock-in allows to detect a spin dependent change with maximum sensitivity while the oscilloscope is used to extract the dc offset as a baseline.
In particular, we use the oscilloscopes mean value within a \SI{0.5}{\second} integration window to get the dc offset for each magnetic field point. We then take the mean value of these points as dc offset. In this recorded data, the PDMR amplitude is also contained within the data for on-resonance points.
As the PDMR and ODMR amplitudes are very small compared to the dc offset ($\approx$4 orders of magnitude), the contribution is negligible.
The same argument holds for the difference between definitions in Eq.~\ref{eqn:contrast_odmr} and Eq.~\ref{eqn:contrast_pdmr}.
Thus in case of low relative amplitudes, our extended definition of contrast is comparable to prior work.

Next, we analyze the dependence of contrast and SNR on the experimental conditions.
To correct for differences in measurement time we normalize the SNR to $t_{\textnormal{norm}}=$~\SI{3600}{\second}.
The value for this time-normalized $\textnormal{SNR}_\textnormal{norm}$ is then calculated by
\begin{equation}
\label{eqn:snr}
 \textnormal{SNR}_\textnormal{norm}=\textnormal{SNR} \sqrt{t_{\textnormal{norm}}/t_{\textnormal{meas}}}
\end{equation}
where $t_{\textnormal{meas}}$ is the total measurement time and SNR the signal-to-noise ratio calculated by dividing the fitted amplitude by the obtained standard deviation noise at  (see Fig.~\ref{fig:magn_field_sens} and calculation of dc magnetic field sensitivity).

\begin{figure}[H]
\centering
\includegraphics[width=1\linewidth]{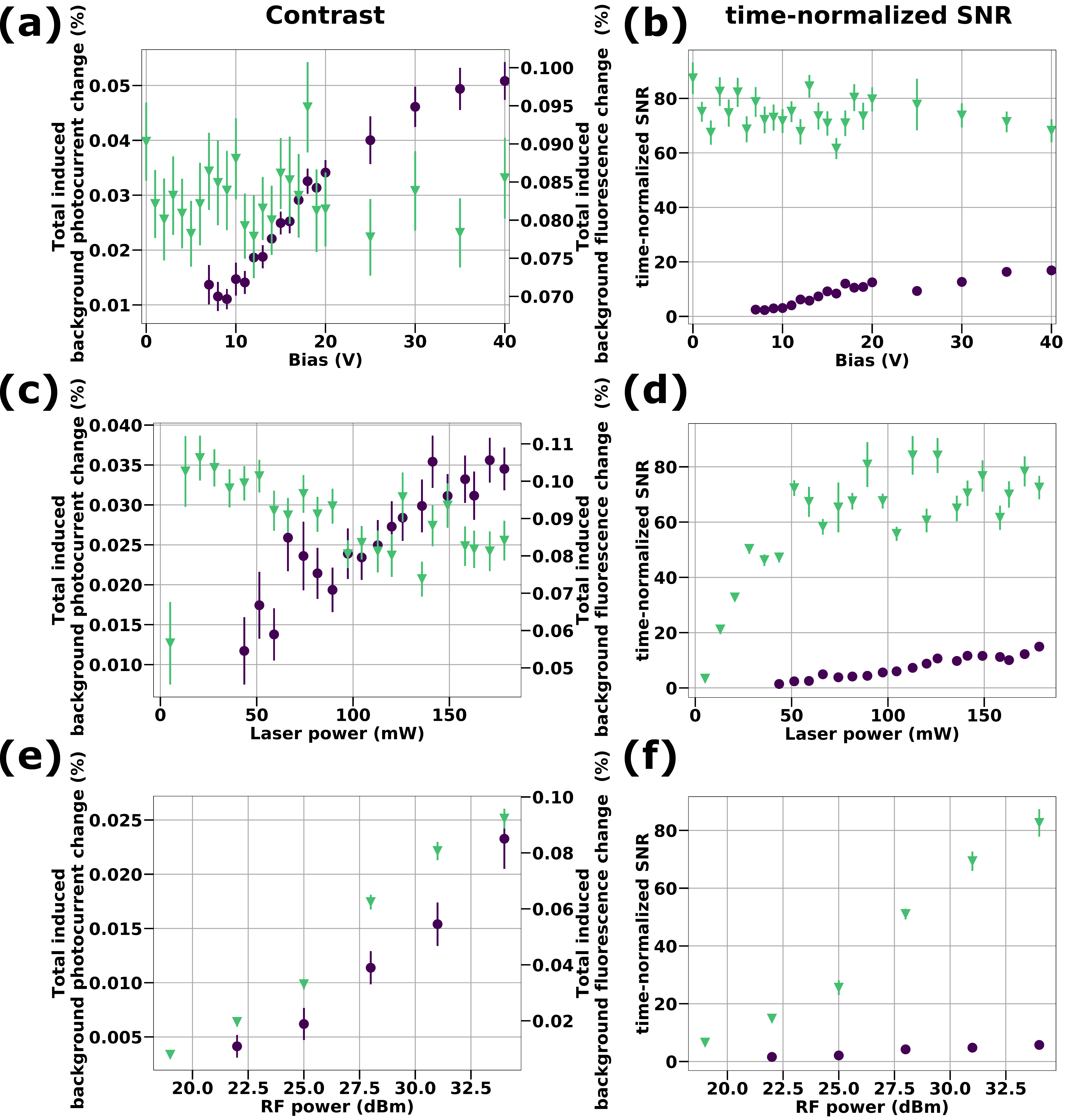}
\caption{\label{fig:contrast} DC referred contrast and time-normalized (\SI{3600}{\second}) SNR dependent on measurement conditions for ODMR (green) and PDMR (purple). Meausurement conditions: Bias $+$\SI{20}{\volt}, \SI{178.5}{\milli\watt} laser and \SI{34}{\dBm} RF power, \SI{223}{\mega\hertz} RF frequency.}
\end{figure}

\begin{figure}[H]
\centering
\includegraphics[width=0.75\linewidth]{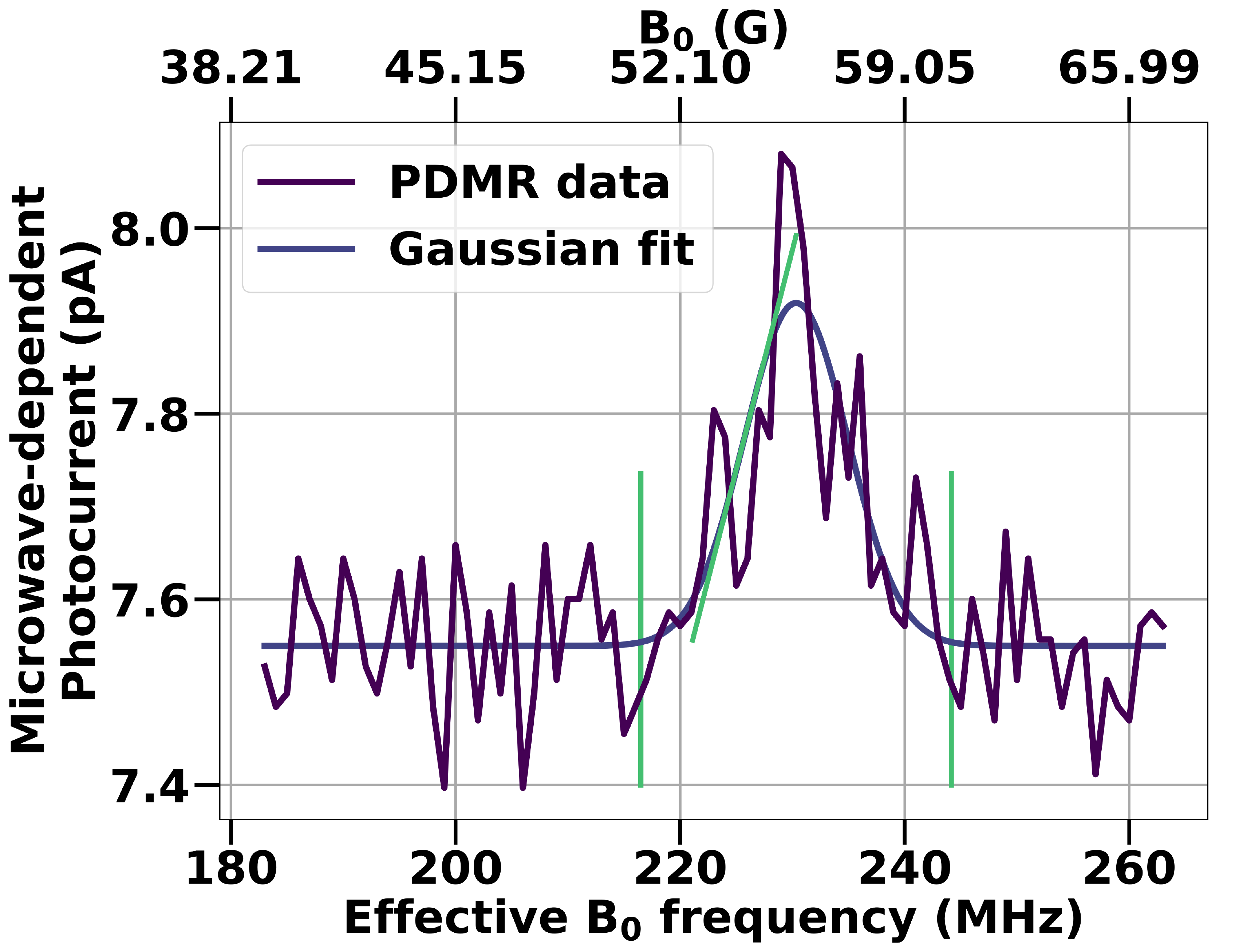}
\caption{\label{fig:magn_field_sens} Extraction of magnetic field sensitivity and SNR for PDMR. The slope is extracted from the fitted Gaussian at $\sigma$ distance from the resonance peak. Noise is determined as standard deviation of measurement points outside of area marked with vertical green lines.}
\end{figure}

In ODMR measurements, the contrast and $\textnormal{SNR}_\textnormal{norm}$ do not depend on the bias, while they do for PDMR  as shown in Fig.~\ref{fig:contrast}(a) and (b).
In PDMR, both contrast and $\textnormal{SNR}_\textnormal{norm}$, are improved for larger bias voltages. 
We attribute this to a better extraction efficiency of free electrons and holes in case of PDMR. 
In terms of contrast a saturating behavior is visible for larger biases. 
As can be seen in Fig.~\ref{fig:parameters}(a), the amplitude is still increasing, thus the dc offset must increase more quickly then the signal in this regime. 
In case of laser power dependence, we see that ODMR contrast decreases for high laser powers, while the PDMR contrast still increases (see Fig.~\ref{fig:contrast}(c)).
The time-normalized SNR shown in Fig.~\ref{fig:contrast}(d) saturates for ODMR, whereas SNR in the PDMR case still increases with  laser power.
If we vary the applied RF power, a clear rise in contrast is visible in both measurement techniques, as depicted in Fig.~\ref{fig:contrast}(e).
$\textnormal{SNR}_\textnormal{norm}$ is increasing for both PDMR and ODMR with applied RF power.
However, larger RF power leads to larger noise for PDMR due to the RF coupling.
The gain in $\textnormal{SNR}_\textnormal{norm}$ is thus bigger for ODMR than for PDMR.

In the following, we calculate the dc magnetic field sensitivity.
For this we use the ODMR and PDMR data shown in Fig.~3(c) in the main text.

The sensitivity is given by comparing the signal power to the noise spectral power.
We estimate the noise by using data points at least 3$\sigma$ apart from the resonance (see Fig.~\ref{fig:magn_field_sens}) and calculating the standard deviation of these data points.
This way, we extract a noise level of \SI{70}{\femto\ampere}.
Signal-to-noise ratio is then obtained by dividing the PDMR resonance amplitude by the noise level. 
The measurement time per point is \SI{25}{\second} for this PDMR measurement, resulting in a noise spectral density of \SI[mode=math]{350}{\femto\ampere/\sqrt{\hertz}}.
The slope of the Gaussian peak is maximum at $\sigma$ distance, related to the FWHM by $\textnormal{FWHM}=2 \sqrt{2\ln{2}}\sigma$ (see Fig.~\ref{fig:magn_field_sens}).
Thus the position of the steepest slope can be found by the amplitude and FWHM of the fitted peak and we find a maximum slope of \SI[mode=math]{1.3}{\nano\ampere/\tesla}.
From this, we calculate a magnetic field sensitivity of \SI[mode=math]{253}{\micro\tesla/\sqrt{\hertz}}.

For the ODMR, we find a noise level of \SI{290}{\micro\volt} within a measurement time of \SI{2.8}{\second} per point. With a slope of \SI{21}{\milli\volt/\tesla} we calculate the sensitivity to be \SI{23}{\micro\tesla/\sqrt{\hertz}}.

Note that we have not reached the saturation of PDMR signal because of the limitation of the laser power. 
Bias and RF power dependence also promise further improvement in SNR and sensitivity.
In addition, while the PDMR contrast is $\approx$1/10 of the ODMR contrast, in theory, comparable values might be achievable, as the underlying ISC process is the same.
Thus the SNR, sensitivities and contrast given in the main text and the supporting information have to be understood as a lower achievable limit.

\subsection{Local magnetic fields variations inside the device}
\begin{figure}[H]
\centering
\includegraphics[width=0.75\linewidth]{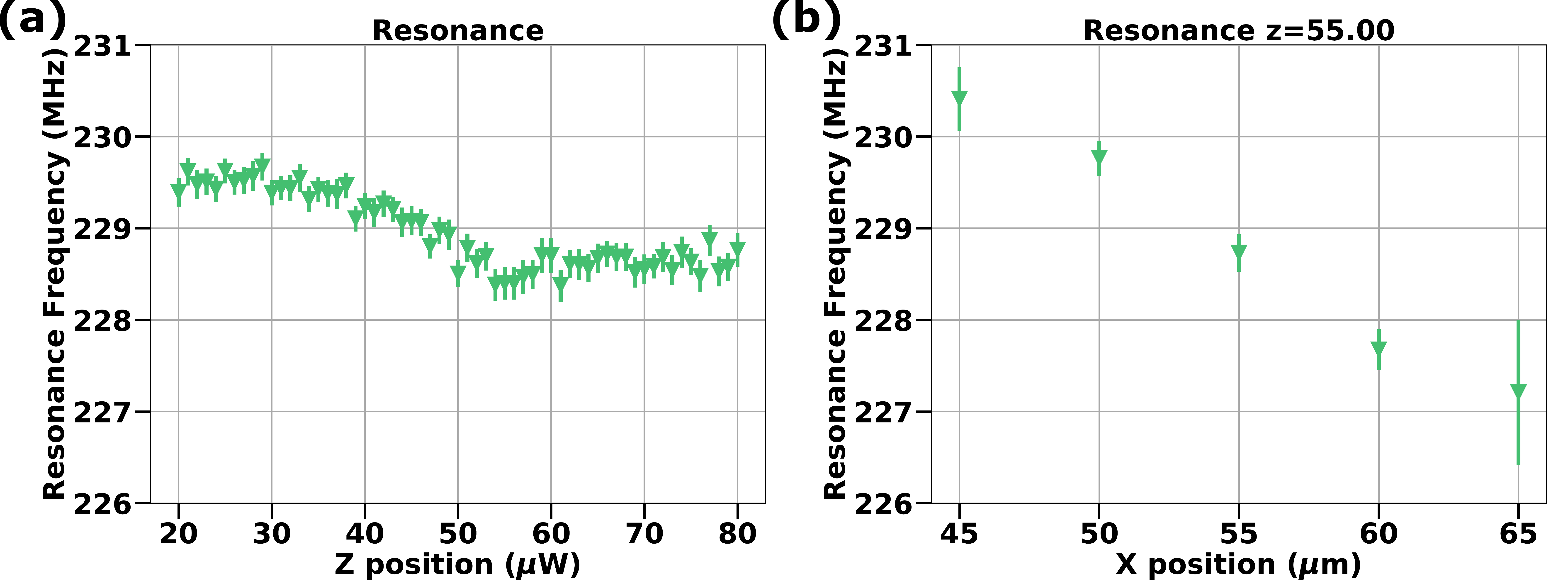}
\caption{\label{fig:magnetization} Magnetic field variation within device. Shift in ODMR resonance peak position depending on (a): $z$-position and (b): $x$-position. }
\end{figure}

In order to check for magnetic fields within the device likely introduced by magnetization of the Ni contacts, we perform ODMR measurements at different depths and x-positions and extract the resonance frequency. 
The results are shown in Fig.~\ref{fig:magnetization}.
As the lines can shift a couple of \si{\mega\hertz}, magnetic field offsets of a few Gauss appear to be plausible in the device structure.
This position-dependent magnetic field might cause the difference in the external fields between PDMR and ODMR because there might be a slight difference in detection volume position.

\bibliography{supplementary_arxiv}
%\end{document}

\end{document}